\documentstyle[mprocl]{article}
 
\def\be{\begin{equation}}
\def\ee{\end{equation}}
\def\bea{\begin{eqnarray}}
\def\eea{\end{eqnarray}}
 
\newcommand{\bfG}{\bf{G}}
\newcommand{\bfe}{\bf{e}}
\newcommand{\bfL}{\bf{L}}
\newcommand{\k}{k}
\newcommand{\calph}{\gamma}

\begin{document}
 
\title{NONHOLONOMIC APPROACH TO ROTATING MATTER IN GENERAL RELATIVITY}

\author{ Mattias Marklund }
 
\address{Department of Plasma Physics, \\
Ume{\aa} University, S-901 87 Ume{\aa}, Sweden \\
e-mail:mattias.marklund@physics.umu.se }
 
\author{ Gyula Fodor, Zolt{\'a}n Perj{\'e}s }
 
\address{KFKI Research Institute for Particle
and Nuclear Physics,\\Budapest 114, P.O.Box 49, H-1525
Hungary \\
e-mail: gfodor@rmki.kfki.hu \quad perjes@rmki.kfki.hu}
 
\maketitle\abstracts{
 Rigidly rotating stationary matter in general relativity has been
 investigated by Kramer (Class.\ Quantum Grav.\ {\bf2} L135 (1985))
 by the Ernst coordinate method. A weakness of this approach is that
 the Ernst potential does not exist for differential rotation. We now
 generalize the techniques by the use of a nonholonomic and
 nonrigid frame. We apply these techniques for differentially rotating
 perfect fluids. We construct a complex analytic tensor,
 characterizing the class of matter states in which both the interior
 Schwarzschild and the Kerr solution are contained. We derive
 consistency relations for this class of perfect fluids.
 We investigate incompressible fluids characterized by these tensors.
}
 
\section{Introduction}
 
  In this contribution, we briefly describe a new approach to rotating
matter in general relativity with applications to differentially
rotating perfect fluids. The full details of this method can be found in
Ref.\ 1.
 
   The essence of our treatment is the generalization of the
Ernst-potential coordinate method to space-times in which the Ernst
potential does not exist at all. (Examples of space-times where the
Ernst
potential does exist are the stationary axisymmetric
vacuum and rigidly rotating perfect fluids.)
   The complex 1-form ${\bfG}$, introduced in Ref.\ 2 exists
whenever a (non-null) Killing vector is given. By its use, we define
a complex nonholonomic basis for axistationary space-times with the
property that it becomes a natural basis whenever an Ernst potential
exists. If the latter is not the case, the basis is noncommutative,
{\em i.e.}\ the structure functions do not vanish.
 
 The 3-space is conformally flat if the Cotton-York tensor
$
  Y_i^{\ell}\ {=}\
  {\epsilon^{jk\ell}}\left(R_{i[j;k]} \right.$ $-\left.
  (1/4)g_{i[j}R_{;k]}\right)
$
vanishes. The {\em Simon tensor}\,\cite{Simon} is essentially an
analytic
continuation
of the Cotton-York tensor in terms of the Ernst potential. Space-times
like the Kerr metric, the interior Schwarzschild, Wahlquist\cite{Wahl}
and the Kramer metrics\cite{Kramer} have a vanishing Simon tensor.
We further generalize the Simon tensor for situations with no Ernst
potential. Solving our field equations for perfect fluids with
a vanishing complex tensor $S_{ik}$, we establish the
following:\vspace{1mm}
 
 {\bf Theorem.}\ {\sl There are no incompressible perfect fluids with a
vanishing $S_{ik}$ tensor.}

\section{Stationary perfect fluid space-times}
 
 We write the metric of a stationary space-time in terms of the metric
$g$ of the Killing trajectories:
\be
  {\rm d} s^2=r({\rm d}t + \omega_i{\rm d}x^i)^2 -
  \frac1r g_{ij}{\rm d}x^i{\rm d}x^j \ .
\ee
Introducing the 3-dimensional complex 1-form
$
{\bf G}\ {\stackrel{\rm def}{=}}\
\left({\rm d}r + {\rm i}r^2\,*\!{\rm d}{\omega}\right)/(2r) ,
$
the Einstein equations become (For our notation, {\em cf.}\ Ref.\ 1):
\begin{eqnarray}
  {G^{ i}}_{;i} & = &({\bfG}\cdot{\bfG}) - ({\bfG}\cdot\bar{\bfG}) +
  {\k}r^{-2}T^*_{oo} , \label{scalar}  \\
  G_{ i;j}-G_{ j;i} & = & \bar G_{ i}G_{ j} - G_{ i}\bar
  G_{ j} + {\rm i}{\k}r^{-2}{\epsilon}_{ ijk}T^{* k}_o ,\\
  R_{ ij} & = & -G_{ i}\bar G_{ j} - \bar G_{ i}G_{ j} -
  {\k}r^{-2}\left(T^*_{ ij}-g_{ ij}T^{*}_{oo}\right) ,
\end{eqnarray}
where $T^{*}_{\mu\nu} {\stackrel{\rm def}{=}} T_{\mu\nu} -
\frac{1}{2}g_{\mu\nu}T$.
The energy-momentum tensor is
$
  T_{\mu}\!^{\nu}=(\mu+p)u_{\mu}u^{\nu}-p\delta_{\mu}^{\nu}\
$
with the normalization condition for the 4-velocity
$
  u_{o}^2-g_{jk}u^j u^k=r \ .
$
 
  We introduce the
complex nonholonomic basis for axistationary space-times
$
 ({{\bfe}}_{\bf 1},{{\bfe}}_{\bf 2},{{\bfL}} ) ,
$
where ${{\bfL}} =\partial/\partial{\varphi}$ is the axial Killing
vector, and the vectors ${{\bfe}}_{\bf 1}$ and ${{\bfe}}_{\bf 2}$
are defined by
\be
       {{\bfG}}=\frac{1}{2r}({\alpha} {{\bfe}}_{\bf 1} +
       {\beta}{{\bfe}}_{\bf 2}) , \quad
  \bar {{\bfG}}=\frac{1}{2r}({\beta} {{\bfe}}_{\bf 1} +
       {\calph}{{\bfe}}_{\bf 2}) ,
\ee
with
$
  {\alpha}=4r^2({\bfG}\cdot{\bfG}),\
  {\beta}=4r^2({\bfG}\cdot\bar{\bfG})
  ,\ {\gamma}=\bar{\alpha} \ .
$
The metric reads
\be
  [{g^{\bf ik}}]= \left[
        \begin{array}{ccc}
        {\alpha}   &   {\beta}  &   0 \\
        {\beta}   &   {{\calph}}  &   0 \\
         0    &    0   &   \varrho^{-2}
        \end{array}\right]  ,
\ee
where $\varrho^2=({\bfL}\cdot{\bfL})$.
From the definition of ${\bfG}$ we have
$r\left({\bfG}+\bar{\bfG}\right)= {\rm d}r$, and
$
  r_{,\bf{1}} \equiv {{\bfe}}_{\bf{1}}r  =
  r_{,\bf{2}} \equiv {{\bfe}}_{\bf{2}}r  = 1/2\ .
$
  The {\em structure functions} ${c^{\bf i}}_{\bf jk}$ are defined
by $[{\bfe}_{\bf j},{\bfe}_{\bf k}]
= {c^{\bf i}}_{\bf jk}{\bfe}_{\bf i}$.
The nonvanishing components are
$
 {c^{\bf{1}}}_{\bf{12}}
 =-{c^{\bf{2}}}_{\bf{12}}
 =2{\rm i}{\k}\varrho T^{*{\varphi}}_o/(r\sqrt{D})
 \ {\stackrel{\rm def}{=}}\ {\varepsilon},
$
where $D \ {\stackrel{\rm def}{=}}\
{\alpha}{{\calph}}-{\beta}^2 < 0$.
The Einstein equation (\ref{scalar}) takes the form
\be
 {\alpha}_{,\bf{1}}+{\beta}_{,\bf{2}}
 - \frac{\alpha}{r} +
 ({\alpha} {{\bfe}}_{\bf{1}}
 + {\beta} {{\bfe}}_{\bf{2}})
 \ln\left(\frac{\varrho}{\sqrt{D}}\right)
 -  \frac{2{\k}}{r}T^*_{oo}
 + ({\alpha}+{\beta}){\varepsilon}=0 \label{Ernst}\ ,
\ee
which reduces to the Ernst equation in vacuum.

\section{The complex tensor}
 
We introduce the {\em complex} tensor
\be
  S_i^{\ell}={\epsilon^{jk\ell}}\left\{
            2g_{ij}g^{rs}G_{[k;|r|}G_{s]}-2G_{k;i}G_{j}
           -{\rm i}{\k}r^{-2}\epsilon_{jk}^{\ \ r}G_{(i}T^*_{or)}
                        \right\} ,
\ee
which is symmetric and trace-free even in the presence of matter.
In vacuo, $S_i^{\ell}$ equals the Simon tensor\cite{Simon}.
 
In the axisymmetric case the condition $S_i^{\ell}=0$ gives two
equations
\be
\alpha_{,{\bf{2}}} = 0 \ , \quad
  ({\alpha}{{\bfe}}_{\bf{1}}+
  {\beta}{{\bfe}}_{\bf{2}})\ln{\varrho}
  -\frac{\alpha}{2r} + \frac14\alpha_{,\bf{1}}
  - \frac{\k}{r}T^*_{oo} = 0 .
\ee
Combining with Eq.\ (\ref{Ernst}), we obtain
\be
  D(\ln\varrho)_{,\bf{1}} +
  \frac{1}{2}{\beta}{{\calph}}_{,\bf{2}} -
  {{\calph}}{\beta}_{,\bf{2}}
  - D{\varepsilon} = 0 \ , \quad
  {\alpha}(\ln c)_{,\bf{1}}=F \ ,
\ee
where
$
 F \ {\stackrel{\rm def}{=}}\  (\k/r)T^*_{oo}-{\beta}{\varepsilon}\
$ and
$
 c \ {\stackrel{\rm def}{=}} \
 D^{-1/2}r^{-1}({\alpha}{{\calph}})^{3/4}\varrho \ .
$
We describe the matter by $\varepsilon$ and $F$.
By the normalization condition of $u^{\mu}$,
\be
  D{\varepsilon}^2+[2(F+\beta\varepsilon)+\k(\mu-p)]
  [2(F+\beta\varepsilon)-\k(\mu+3p)]=0\ .
\ee
The Bianchi identities give
\be
  {\k}p_{,\bf{1}} = \frac1{2r}\left\{{\k}p + \frac12{\varepsilon}
    ({\beta}-{{\calph}}) - F - \beta\varepsilon
  + [2r(F + \beta\varepsilon)
           - {\k}(\mu+3p)r](\ln\varrho)_{,\bf{1}}\right\} .
\ee
The remaining field equations are as follows,
\begin{eqnarray}
  {\alpha}{\alpha}_{,\bf{1}\bf{1}}-\frac{3}{4}
({\alpha}_{,\bf{1}})^2 & = &
  4{\alpha}( 2{\varepsilon}{\beta}_{,\bf{1}} +
  {\beta}{\varepsilon}_{,\bf{1}} + F_{,\bf{1}})
- 2(3{\beta}{\varepsilon} +
  2F){\alpha}_{,\bf{1}}
  \nonumber\\
  &&+ 4F(2{\beta}{\varepsilon} + F) +
  \frac{4\alpha}{r}\left({\beta}{\varepsilon}
  + r{{\calph}}{\varepsilon}^2 + F - {\k}p\right) \label{eik} \\
  F_{,\bf{1}} + {\varepsilon}F&=&{{\calph}}{\varepsilon}^2 +
  \frac{\calph}{2rD}\left[{\k}(\mu+3p){\beta} - \left({\alpha}{{\calph}}
  +{\beta}^2\right){\varepsilon}-2F{\beta}\right]\\
  {\alpha} F_{,\bf{1}}+({\alpha}-{\beta}){\varepsilon} F
  &=&{\alpha}{\varepsilon}({{\calph}}-{\beta})\left({\varepsilon} -
  \frac1{2r}\right)
                -\frac34{\beta}{\varepsilon}{\alpha}_{,\bf{1}}
                -{\alpha}{\beta}{\varepsilon}_{,\bf{1}}\ .
\end{eqnarray}
In general, the first order equations can be solved for the
$\bfe_{\bf 1}$ derivatives of all our functions except for the fluid
energy density $\mu$. For an
incompressible fluid the second order equation (\ref{eik}) together
with the integrability conditions of our variables gives a system of
algebraic equations.
The proof of our Theorem follows by showing the inconsistency of these
equations. If $\mu$ is not constant, one can still obtain equations
linear in $\mu_{\bf,1}$ and $\mu_{\bf,2}$. At this time we do not
know if the resulting algebraic equations
yield any other solution than those found by
Wahlquist\cite{Wahl} and Kramer\cite{Kramer}.

\end{document}